\documentstyle[aps,prb,multicol,graphics]{revtex}

\newcommand{\be}{\begin{equation}}
\newcommand{\ee}{\end{equation}}
\newcommand{\bea}{\begin{eqnarray}}
\newcommand{\eea}{\end{eqnarray}}
\newcommand{\Lrule}{\vspace*{-0.2in}\noindent\vrule width3.5in height.2pt
  depth.2pt \vrule depth0em height1em}
\newcommand{\Rrule}{\vspace{-0.1in}\hfill\vrule depth1em height0pt \vrule
  width3.5in height.2pt depth.2pt\vspace*{-0.1in}}
\renewcommand{\narrowtext}{\begin{multicols}{2} \global\columnwidth20.5pc}
\renewcommand{\widetext}{\end{multicols} \global\columnwidth42.5pc}

\newcommand{\etal}{{\it et al.}\ }

\begin{document}

\title{Striped States in Quantum Hall Effect: \\ Deriving a Low Energy Theory from Hartree-Fock}

\author {Anna Lopatnikova${}^{(1,2)}$, Steven H.~Simon${}^{(1)}$, Bertrand I.~Halperin${}^{(2)}$, and Xiao-Gang Wen${}^{(3)}$}
 
\address{${}^{(1)}$ Lucent Technologies Bell Labs, Murray Hill, New Jersey 07974 \\ ${}^{(2)}$ Department of Physics, Harvard University, Cambridge, Massachusetts 02138\\
 ${}^{(3)}$ Department of Physics, Massachusetts Institute of Technology, Cambridge, Massachusetts 02139}

\maketitle

\begin{abstract}

There is growing experimental and theoretical evidence that very clean two dimensional electron systems form unidirectional charge density waves (UCDW) or ``striped'' states at low temperatures and at Landau level filling fractions of the form $\nu = M + x$ with $4 < M < 10$ an integer and $0.4 \lesssim x \lesssim 0.6$.  
Following previous work, we model the striped state using a Hartree Fock approach.  
We construct the low energy excitations of the system by making smooth deformations of the stripe edges analogous to the construction of edge state excitations of quantum Hall droplets.  
These low energy excitations are described as a coupled Luttinger liquid theory, as discussed previously by MacDonald and Fisher (Phys.~Rev.~B {\bf 61}, 5724 (2000)).  
Here, we extend that work and explicitly derive all of the parameters of this low energy theory using a Hartree Fock approach. 
We also make contact with the equivalent low energy hydrodynamic approach of Fogler and Vinokur (Phys.~Rev.~Lett.~{\bf 84}, 5828 (2000)) and similarly derive the parameters of this theory.
As examples of the use of these results, we explicitly calculate the low-energy excitation spectrum and study tunneling into the striped state.

\end{abstract}

\begin{multicols}{2}

\section{Introduction}

Even before the discovery of the quantum Hall effect, it was suggested that the ground state of a partially filled Landau level of a two dimensional electron gas (2DEG) in a high magnetic field should be a charge density wave\cite{fukuyama}. 
It was clear that (neglecting disorder) due to the degeneracy of the noninteracting electron spectrum, the ground state of such a partially filled Landau level should be determined entirely by electron-electron interactions, and charge density wave states were obvious candidates.  
In fact, when the fractional quantum Hall effect was discovered for the one third filled Landau level, it was initially identified as a charge-density wave phenomenon.
However, it was quickly determined that such observed fractional quantum Hall states in the lowest Landau level are uniform density fluids.  (States that appear to be Wigner crystals were later observed in the lowest Landau level at still higher fields\cite{shayegan}).

In partially filled higher Landau levels, however, there is now growing experimental\cite{experiment1,stormer} and theoretical\cite{fogler1,moessner,macdonald1,fertig1,haldane} evidence that charge density wave states do occur at low temperatures in ultra-clean samples.  
The experimental phenomenology of these high Landau level states, which has been elucidated by a series of publications by two groups\cite{experiment1,stormer}, is quite rich and complex.  
For filling fractions $\nu = n \phi_0/B$ (with $n$ the electron density, $B$ the magnetic field, and $\phi_0 = h c/e$ the flux quantum) between roughly 9/2 and 21/2, certain trends have been observed that may indeed be associated with charge density wave behavior.  
In particular, for filling fractions of the form $\nu = M + x$ with $M$ an integer between roughly 4 and 10 and $ 0.4 \lesssim x \lesssim 0.6$ an anisotropic state is observed that has been tentatively identified as a unidirectional charge density wave (UDCW) or ``striped'' state.

Although the discovery of this state attracted quite a bit of attention in the community, perhaps it should not strike us as surprising.   Indeed, such a state had been predicted for the higher Landau levels several years earlier\cite{comment} by Koulakov, Fogler, and Shklovskii\cite{fogler1} and by Moessner and Chalker\cite{moessner}. Koulakov \etal found that, within the Hartree-Fock approximation, when the partially filled Landau level is nearly half-filled, the ground state of a 2DEG is a UCDW that looks like a set of stripes in momentum space. The filling factor in the stripes alternates between $[\nu ]$ and $[\nu ]+1$ ($[X]$ is the integer part of $X$) with a period, $a$, about 2.7 times the cyclotron radius $R_c$ ($R_c = \sqrt{2L+1}\, l$, where $L$ is the partially filled Landau level, $L=[M/2]$, and $l$ is the magnetic length, $l=(\hbar c/eB)^{1/2}$).
Moessner and Chalker confirmed the result for finite temperatures and justified the use of the Hartree-Fock approximation by proving it to be exact in the limit of infinite Landau level for the fermion hard-core potential (the potential that gives the exact Laughlin wavefunction at the lowest Landau level). 
More recently, extensive exact diagonalization studies on small systems have provided strong evidence that the ground state of a partially filled higher Landau level is indeed a charge density wave\cite{haldane}.
The UCDW state is anisotropic and it is feasible that its low energy transport properties are anisotropic as well.

There have been suggestions, however, that, for a Coulomb interaction, the ground state is a more complicated structure than a UCDW.   A renormalization-group analysis by MacDonald and Fisher (MF) suggests 
an instability of the UCDW against formation of a (highly anisotropic) Wigner crystal in the presence of backscattering\cite{macdonald1}. Crystallization was also found in numerical Hartree-Fock calculations of C\^{o}t\'{e} and Fertig\cite{fertig1}. 
However, C\^{o}t\'{e} and Fertig found the interactions of the modulations among different stripes to be so weak as to render the effects of the Wigner-crystal instability unmeasurable.  
The long wavelength behavior of the anisotropic crystalline state, derived in that work\cite{fertig1}, is indistinguishable from that of a UCDW state, and, for practical purposes, one can ignore backscattering as negligibly small and consider the ground state to be a UCDW.  
Thus, throughout this paper we will make the assumption that the ground state is well described as a UCDW. 

To go beyond the ground state properties one needs a way of describing excitations around the ground state and effective interactions between them.  This is the main objective of this paper. The low energy excitations can be described as smooth ``elastic'' deformations of the UCDW ground state.   We represent these excitations as coherent states and derive an effective interacting Luttinger liquid model for them, complete with analytical expressions for the interaction matrix, on the basis of a first-principles Hartree-Fock calculation.

The interaction matrix of the Luttinger liquid model, which we calculate, can be expanded for long wavelengths to get what essentially are the low energy elastic parameters of the system.  These quantities can be used as inputs for calculating a great number of physical quantities.  
In Ref.~\onlinecite{fogler2} the elastic parameters were used as an input for a calculation of transport properties of the system.  In Ref.~\onlinecite{macdonald1}, MF use these parameters as inputs 
for the renormalization-group analysis of backscattering. 
The parameters are also used in Ref.~\onlinecite{wexler1} to study the Kosterlitz-Thouless like dislocation unbinding transition of the striped phase.  The calculation of the elastic parameters in that reference agrees with our work in the long wavelength limit\cite{wexler1}. (Our work, however, is also valid at higher wavevector). 
Similarly, in  Ref.~\onlinecite{scheidl} these parameters are used as inputs for an RG calculation of the effects of disorder on the striped state\cite{scheidl}. 
Finally, in Sec.~\ref{sec:dynamics} and \ref{sec:tunneling} of this paper, we will 
use these parameters in calculations of the dynamical properties and the I-V spectrum for tunneling of electrons into a striped state.   We certainly expect that many more applications of these results will be found. 

In Sec. \ref{sec:lowenergy} we will outline the expected form of the low energy Luttinger liquid theory -- including a detailed discussion of symmetry considerations.  We will also make an explicit connection between the form of the Luttinger liquid theory proposed by MF\cite{macdonald1} and the low energy phenomenological theory proposed in Ref.~\onlinecite{fogler2}.  

Section~\ref{sec:excitations} will be devoted to the description of the low energy excitations in terms of semi-classical coherent state operators.  We begin by discussing the Hartree-Fock ground striped ground state.  Deformations of this ground state are described as coherent excitations of elementary density operators. 
We will show how our coherent state description naturally leads to a quantized Luttinger liquid Hamiltonian. 

In Sec.~\ref{sec:parameters} we will calculate the energies of low energy coherent state excitations and hence obtain the full interaction matrix of the Luttinger liquid theory.  In Sec.~\ref{sec:results}, we expand the interaction matrix for long wavelengths to find the parameters of the low energy theory of Ref.~\onlinecite{fogler2}.

Sections~\ref{sec:dynamics} and \ref{sec:tunneling} will give examples of applications of the Luttinger liquid theory and its parameters to the dynamics of and tunneling into the striped state.

\section{Form of The Low Energy Theory}
\label{sec:lowenergy}

In this paper, we assume that the UCDW ground state is a set of straight electron (or hole) density stripes, continuous in the $x$-direction\cite{macdonald1}. 
Throughout the paper, each stripe is labeled with integer indices $I$ or $J$. The edges of the stripes are labeled $\alpha $ or $\beta $ that take the values $R$ (or ``+1'') for the right edge and $L$ (or ``-1'') for the left edge. 
The requirement that the UCDW state minimize the energy fixes the distance between the centers of the stripes, $a$.
All our microscopic analysis is restricted to the partially filled Landau level.
The only contribution of the lower, completely filled, Landau levels to our model is screening of the Coulomb interactions\cite{aleiner1}. 

The general form of the Luttinger liquid model has been written down by MF on phenomenological grounds\cite{macdonald1}.
In their work, MF assume that the low energy excitations are small displacements of the edges of the stripes.
The dynamics of the edges is represented by a quadratic Hamiltonian for the displacement fields, $u_{\alpha}({\bf q})$.
MF argue that the displacement fields are associated with the chiral currents on stripe edges, $\rho_{\alpha }({\bf q})$,  as $u_{\alpha }({\bf q}) = \alpha 2\pi l^2\rho_{\alpha }({\bf q})$, and the Hamiltonian for the currents has a form very similar to that of the displacement fields\cite{fourier}:
\be
	H = \frac{l^2}{2 a^2}\int d{\bf q} \sum_{\alpha ,\beta }\alpha \beta \rho_{\alpha ,{\bf q}} D_{\alpha \beta } (-{\bf q}) \rho_{\beta ,-{\bf q}}.\label{eq:hamilt-mf}
\ee
The interval of integration over $q_x$ is cut off by $2\pi /l$;  $q_y$ is integrated over the interval $[-\pi /a;\pi /a]$; $D_{\alpha \beta } ({\bf q})$ is the effective interaction matrix.
Except for a prefactor of $1/a$, this is exactly the Hamiltonian given by MF.  
(The prefactor appears because of the difference between our Fourier transform conventions.)
This classical theory is quantized by imposing Ka\v{c}-Moody commutation relations onto the currents $\rho_{I\alpha }({\bf q})$. (See Sec.~\ref{subsec:rho} below.) 

As MF point out, the general form of the elements of the interaction matrix, $D_{\alpha \beta } ({\bf q})$, are constrained by the symmetries of the ``striped'' state\cite{macdonald1,fradkin1}.
Symmetry considerations become more obvious if we introduce symmetric ($S$) and anti-symmetric ($A$) modes that for each edge are defined as
\bea
	\rho_{S ,{\bf q}} & = & \rho_{R ,{\bf q}}e^{iq_ya\frac{\nu^*}{2}} - \rho_{L ,{\bf q}}e^{-iq_ya\frac{\nu^*}{2}} \\
	\rho_{A ,{\bf q}} & = & \rho_{R ,{\bf q}}e^{iq_ya\frac{\nu^*}{2}} + \rho_{L ,{\bf q}}e^{-iq_ya\frac{\nu^*}{2}},
\eea
where $\nu^*=\nu-[\nu ]$ is the fractional filling of the partially filled Landau level.
The phases $e^{\pm iq_ya\frac{\nu^*}{2}}$ enforce the particle-hole symmetry of these modes at $\nu = 1/2$.
These factors are also natural in the sense that the right (left) edge is displaced a distance $\pm a \nu^*/2$ from the center of the stripe, and thus an additional phase is included in its $q_y$ Fourier transform.

The Hamiltonian for the symmetrized excitations becomes
\bea
       	H & = & \frac{l^2}{2a^2}\int d{\bf q} \left[ D_{S} (-{\bf q}) |\rho_{S,{\bf q}}|^2 + D_{A} (-{\bf q}) |\rho_{A,{\bf q}}|^2 \right. \nonumber \\
	& & + \left. \mbox{Im}\{ \rho_{A,{\bf q}}\rho_{S,-{\bf q}}\} \mbox{Im} \{e^{iq_ya\nu^*} D_{LR} (-{\bf q})\}  \right], \label{eq:hamilt3}
\eea 
where the symmetric and antisymmetric coupling constants relate to the interaction matrix $D_{\alpha \beta } ({\bf q})$ as
\bea
	D_S({\bf q}) & = & \frac {1}{4} \sum_{\alpha ,\beta} e^{iq_ya\frac{\nu^*}{2}(\alpha -\beta )}D_{\alpha \beta }({\bf q}) \nonumber\\
	D_A({\bf q}) & = & \frac {1}{4} \sum_{\alpha ,\beta} \alpha \beta e^{iq_ya\frac{\nu^*}{2}(\alpha -\beta )}D_{\alpha \beta }({\bf q}).\label{eq:sym-ant}
\eea
This change of variables separates the two components of any excitation around the ``striped'' ground state: the displacement of the guiding center and the change in width of the stripes.

\begin{figure*}
	\centering
	\scalebox{0.5}{\includegraphics{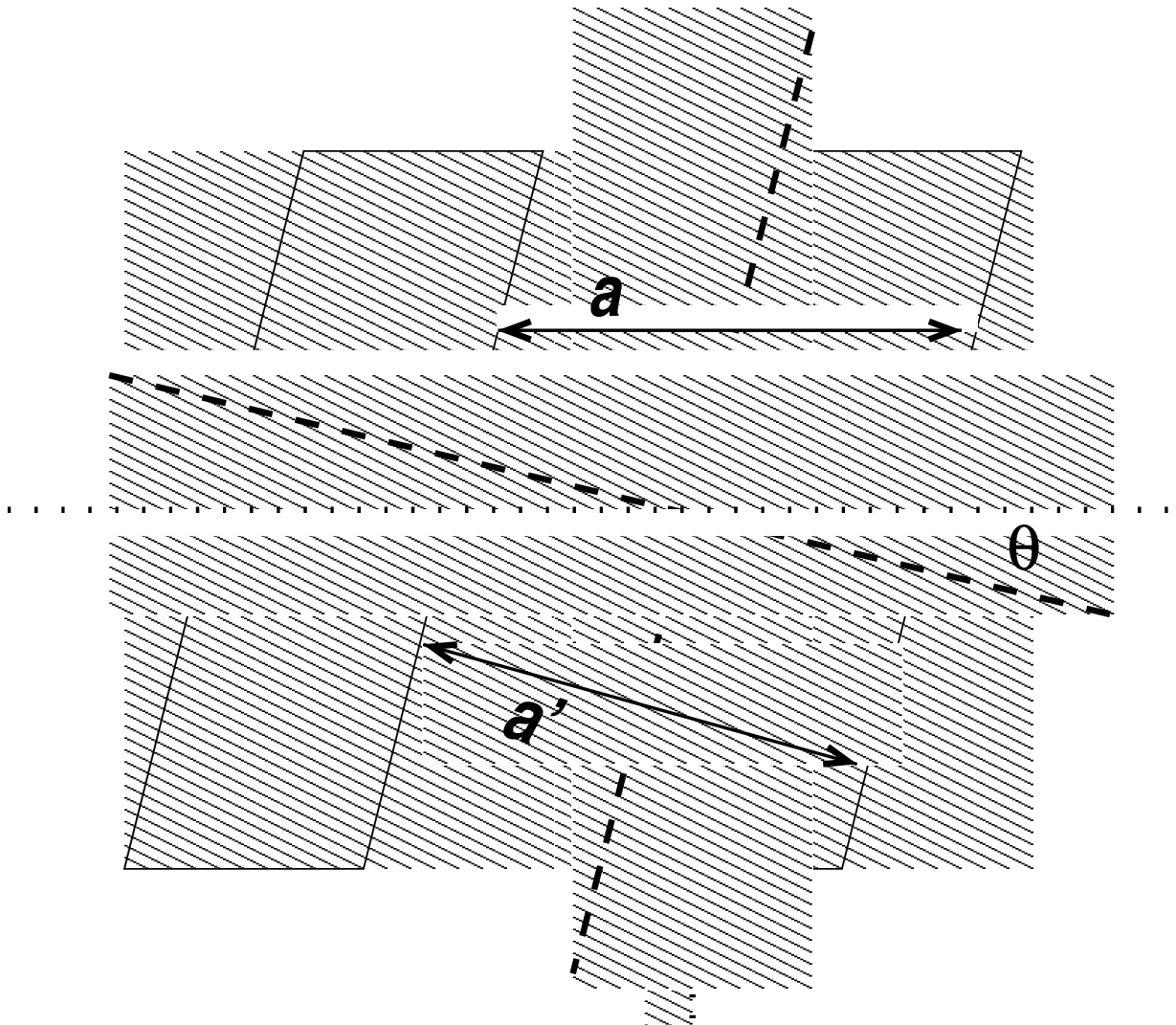}}
	\caption{\label{fig:shear} A long wavelength symmetric excitation locally behaves like a shear rotation.  Such a rotation changes the distance between the stripes, so that the new distance $a'=a \cos \theta \approx a (1-\theta^2/2) $, where $\theta $ is the angle of the local rotation.  The angle $\theta $ is proportional to $q_x$, so that $a-a' \propto q_x^2$.   }
\end{figure*}

The displacement of the guiding center of each stripe is described by the symmetric component, $\rho_{S ,{\bf q}}$.  In this case, the excitations on the opposite edges of each stripe are in phase.
This is equivalent to saying that the width of each stripe stays the same, but the guiding center of the stripe fluctuates around its equilibrium position.   As was mentioned by MF and others, the form of the elasticity for this mode should be like that of smectic liquid crystals, $D_S({\bf q}) \propto c_1 q_y^2+c_2 q_x^4$, {\it i.e.} there is no $q_x^2$ contribution.  
To see why the $q_x^2$ term should be absent, we consider a long wavelength symmetric perturbation in the $q_x$-direction.   
Locally, such an excitation looks like a ``shear rotation'' as shown in Fig.~\ref{fig:shear}.  
Although a uniform rotation costs no energy, this local shear rotation also has the effect of changing the perpendicular spacing of the stripes.   The 
change in spacing of the stripes (for fixed amplitude of perturbation) is proportional to $q_x^2$ (since the angle of the shear rotation with respect to the unperturbed state is proportional to $q_x$).   However, the initial stripe spacing $a$ is already chosen to minimize the ground state energy of the system $E_{GS}$ so that $d E_{GS}/da = 0$.  Thus, the leading term of the energy cost of the shear rotation is proportional to the square of the change in spacing, or as $q_x^4$.     It also must be noted, that the rotational symmetry can be broken by disorder or the geometry of the sample and the $q_x^2$ component may appear, changing various properties of the system, such as RG scaling or tunneling.

The anti-symmetrized field, $\rho_{A,{\bf q}}$, on the other hand, represents the local change in the width of each stripe.  At distances large compared to the period of the UCDW, $a$, the stripe width fluctuations look like an electron density wave, so that the matrix element $D_{A} ({\bf q})$ must be proportional to $V({\bf q})$, the appropriately screened Coulomb interaction.
The leading corrections to this singular Coulomb form of the long wavelength density-density interaction will be called the inverse compressibility $\chi^{-1}$.    This inverse compressibility is simply related to the energy change for changing the density if the Coulomb piece is compensated by a background charge that leaves the system locally neutral on scales larger than the stripe spacing. 

Using only these kind of symmetry arguments and the similarities between the UCDW states of the 2DEG and the smectic liquid crystals, Fogler and Vinokur\cite{fogler2} (FV) proposed an equivalent hydrodynamic Hamiltonian.  (A similar elastic description is written in Ref.~\onlinecite{scheidl}.)
Expanded for long wavelength excitations to second order in the fluctuations, the Hamiltonian of FV may be written as $H_{FV}=H+H_K$, where $H_K$ is the kinetic energy term, and
\bea
	H & = & \frac{1}{2}\int \frac{d{\bf q}}{(2\pi )^2} \left[ (Yq_y^2 + K q_x^4)|u_{\bf q}|^2 \right.  \nonumber \\
	& & \mbox{} + (V({\bf q})+\chi^{-1})|\delta n_{\bf q}|^2 + 2imCq_yu_{\bf q}\delta n_{\bf -q} ]\label{eq:fogler}.
\eea
The hydrodynamic fields are $n_{\bf q}$, the coarse-grained component of electron density, ($mn$ is the mass density) and $u_{\bf q}$, the displacement of the guiding center of each stripe. 
The field $\delta n_{\bf q}$ is the displacement of $n_{\bf q}$ from its equilibrium value $n_0=\frac{\nu^*}{2\pi l^2}$.  
The coefficients $Y$ and $K$ are the compression and bending elastic moduli, $V({\bf q})$ is the appropriately screened Coulomb potential, and $\chi $ is the compressibility.
Here, the inverse compressibility, as discussed above  is just the long wavelength density-density interaction once the Coulomb interaction is ``removed'' by compensating with a neutralizing background charge.    Since the ground state energy is always measured with such a neutralizing background,  the inverse compressibility 
can be written as the change in energy when the stripe width is changed by a small fraction while the distance between the centers of the stripes is kept constant, 
\begin{equation}
\label{eq:chiinv}
\chi^{-1}=2\pi l^2 \left. \frac{\partial^2 E_{GS}}{\partial \nu^2 }\right|_a
\end{equation}
 with $E_{GS}$ the energy of the striped ground state.  
The constant $C$, above, is the measure of the change in the local period of the stripes with the change in local electron density.  The condition of equilibrium derived from the Hamiltonian gives a relation\cite{fogler2} 
\be
\label{eq:C}
	C=-\frac{Y}{ma}\frac{\partial \langle a \rangle }{\partial n_0}.
\ee
This is obtained by requiring that for a given long wavelength variation in $n({\bf r})$, with $q_x = 0$ and $q_y \to 0$, the energy should be minimized with respect to $u$ ({\it i.e.} $\delta H/\delta u = 0$), when the change in the local lattice constant ($\delta a = a \partial_y u$) has the value required by local equilibrium ($\delta a = \delta n \partial \langle a \rangle /\partial n_0$).

Since the field $u_{\bf q}$ in the Hamiltonian is the displacement of the guiding center of each stripe, it corresponds to the symmetrized field $\rho_{S,{\bf q}}$ 
\be
u_{\bf q} =  2\pi l^2\frac{\rho_{S,{\bf q}}}{2}.
\ee
The other field in the Hamiltonian, $\delta n_{\bf q}$, is the deviation of the coarse-grained electron density from the equilibrium value, $n_0$; its counterpart in the Luttinger liquid description is the anti-symmetrized field  $\rho_{A,{\bf q}}$.
Smearing  $\rho_{A,{\bf q}}$ over the stripe width leads to the correspondence 
\be
\delta n_{\bf q} =  \frac{\rho_{A,{\bf q}}}{a}
\ee

Thus, the first three terms of the hydrodynamic Hamiltonian exactly correspond to the three terms in the Luttinger liquid Hamiltonian as represented in Eq.~(\ref{eq:hamilt3}).
The last two terms are introduced in Ref. \onlinecite{fogler2} for bookkeeping purposes only. 

The kinetic energy term $H_K$ disappears when one imposes a strong magnetic 
field and projects onto a single Landau level.  The projection, however, has 
the effect of changing the commutation relations, so that the variables $u$ 
and $\delta n$ no longer commute with each other.  (The changed commutation 
relations were not explicitly stated by FV, but were used implicitly in their 
equations of motion).

FV included an additional counter-term in their Hamiltonian of form 
\be
	\lambda [\partial_y u-\frac{1}{2}|\nabla u|^2 ]
\ee
This counter-term is necessary when terms beyond second order are included in 
the Hamiltonian, in order to assure that rotational symmetry is maintained and 
that $\langle \nabla u\rangle = 0$ in equilibrium.  Here, we only need the Hamiltonian to 
second order, so the counter-term is absent.

For convenience, we adopt the notation of the hydrodynamic Hamiltonian to write down the expected form of the Luttinger liquid interaction matrix ${\bf D}({\bf q})$ in the long wavelength limit
\bea
	D_S({\bf q}) & = & \frac{l^2 a^2}{4} (Y q_y^2 + K q_x^4), \nonumber\\
	D_A({\bf q}) & = & \frac{1}{2\pi }\frac{V (q)+\chi^{-1}}{2\pi l^2}, \nonumber\\
	\mbox{Im}\{ e^{iq_ya\nu^*} D_{RL}({\bf q}) \} & = & -\frac{ma}{2\pi }Cq_y.\label{eq:d-hydr}
\eea

In the course of the paper, we will develop a microscopic Hartree-Fock description of the ``striped'' state and see the Hartree-Fock theory exactly confirm the phenomenological expectations of the form of the Luttinger liquid (and, thus, hydrodynamic) Hamiltonian.
The microscopic description will also give us quantitative results for all the physical parameters of the effective Hamiltonians above as well as the necessary commutation relations to derive the correct dynamics of the system from the effective Hamiltonian (Eq.~\ref{eq:fogler}) of FV or the Luttinger liquid Hamiltonian (Eq~\ref{eq:hamilt-mf}) of MF.

In Section~\ref{sec:excitations} below we show how our coherent state description of low energy excitations naturally leads to a quantized Luttinger liquid Hamiltonian that has the form of Eq.~(\ref{eq:hamilt-mf}) and can, in principle, be used to obtain the anharmonic terms not discussed in this paper.
We will in fact find that the Luttinger liquid Hamiltonian is precisely equivalent to Hartree-Fock within the low energy subspace defined by linear combinations of $\rho_\alpha ({\bf q})|G\rangle $ ($|G\rangle $ is the ground state).

In Sec.~\ref{sec:parameters} we will calculate the energies of low energy coherent-state excitations and, in Sec.~\ref{sec:results}, obtain the coefficients $Y$, $K$, $C$, and $\chi $ using the Hartree-Fock approximation.
This is the main quantitative result of the paper.
A parameter similar to $Y$ has actually been calculated by MF.  

The parameters $Y$, $K$, $C$, and $\chi $ fully define the interaction matrix $D_{\alpha \beta } ({\bf q})$ in the long wavelength limit, as well as the hydrodynamic Hamiltonian (Eq.~(\ref{eq:fogler})).
These parameters have also been calculated in Hartree-Fock by Wexler and Dorsey\cite{wexler1}, and agree with the long wavelength limit of our results.
Our results, however, are more general and are valid at smaller wavelengths.
Thus, we will have presented a step-by-step derivation of the effective interacting Luttinger liquid Hamiltonian with all parameters on the basis of a microscopic Hartree-Fock calculation.

\section{Microscopic Description of Excitations}
\label{sec:excitations}

\subsection{Hartree-Fock Ground State}
\label{subsec:grstate}

To derive a low energy effective theory from a microscopic Hartree-Fock theory, one has to find a way of treating low energy excitations within Hartree-Fock. 
But first, as a reference point, one has to construct a ground state.

A Hartree-Fock state is a single Slater determinant made up of one-electron wavefunctions.  
Since the ground state, $|G_{\nu^*}\rangle $, is assumed to be a UCDW, translationally invariant in the $x$-direction, the one-electron wavefunctions of choice are the solutions to the free-electron Schr\"{o}dinger equation in the Landau gauge, ${\bf A} = (-By,0,0)$:
\bea
	\psi_{L,{k_x}_n} & = & \frac{1}{\sqrt{L_x}} e^{i{k_x}_nx} \phi_L(y-{k_x}_nl^2) \nonumber\\
	\phi_L(y) & = & \frac{1}{\sqrt{l2^LL!\sqrt{\pi}}}e^{-\frac{y^2}{2l^2}}H_L\left( \frac{y}{l} \right) \nonumber,
\eea
where $L_x$ is the length of the system in the $x$-direction (assumed to have periodic boundary conditions), ${k_x}_n = \frac{2\pi}{L_x}n, n=1,2,..,N$, and $N$ is the number of electrons needed to fill a Landau level.
The pseudo-momentum $k_x$ controls the position of the wavefunctions (the wavefunctions are centered around $k_xl^2$ in $y$-direction) and labels the states.
A striped state of non-interacting electrons with filling fraction $\nu^*$ and distance between the stripes $a$ can be created by filling all the wavefunctions with $k_x \in [na/l^2 , (n+\nu^*)a/l^2]$ ($n$ is an integer) and leaving the others empty.

For future use, we construct a vector $\Psi_L $ of the complete set of the one-electron wavefunctions in the $L$th Landau level:
\be
	\Psi_L =  \left[ \begin{array}{c}
				\psi_{{k_x}_1} \\
				\psi_{{k_x}_2} \\
				\vdots \\ 
				\psi_{{k_x}_n} \\
				\vdots \\
 				\psi_{{k_x}_N}
				\end{array} \right] \label{eq:psi}
\ee
We restrict the Hilbert space of the one-electron states to the highest partially-filled Landau level, $L$, assuming that the presence of lower completely filled Landau levels manifests itself only in screening of the Coulomb interaction\cite{aleiner1}.

We can define an anti-symmetrizing operator ${\cal A}$ that, when acting upon  $\Psi_L$, creates a Slater determinant of the set of the one-electron wavefunctions comprising $\Psi_L$.
This Slater determinant would represent a Hartree-Fock wavefunction for a completely filled ($\nu^* = 1 $) Hartree-Fock state.

To represent the striped ground state, $|G_{\nu^*}\rangle $, however, we need to make a Slater determinant of a subset of the one-electron wavefunctions that correspond to the filled states.
By analogy with ${\cal A}$, we can define an operator ${\cal A}_{a,\nu^*}$ that 
creates such a Slater determinant for the striped state with filling factor $\nu^*$ and distance between the stripes $a$.  
When it acts upon $\Psi_L$, the operator ${\cal A}_{a,\nu^*}$ effectively picks out the wave-functions of the filled one-electron states and anti-symmetrizes them.
We use the resulting Slater determinant, ${\cal A}_{a,\nu^*}\Psi_L$, as the Hartree-Fock trial ground state of the striped quantum Hall system.
In the spirit of variational Hartree-Fock, the width of the stripes, $a$, is the parameter to be adjusted by minimizing the Hartree-Fock energy of the state, as will be shown in Sec.~\ref{sec:parameters}.
The Slater determinant ${\cal A}_{a,\nu^*}\Psi_L$ with the optimal distance between the stripes is our Hartree-Fock ground state.

\subsection{Coherent States}
\label{subsec:rho}

Having constructed the striped ground state, we can proceed to describe low energy excitations around it.  
In this section, we will introduce a method for creating any low energy excitation around the striped ground state in a way that connects Luttinger liquids with Hartree-Fock states.

We start from the Luttinger liquid description of excitations.
As was mentioned in the Introduction, we assume that the ground-state UCDW has a striped structure in momentum space, with the filling factors in the stripes alternating between integers $[\nu ]$ and $[\nu ]+1$.
Effectively though, the filling factors in the stripes are $0$ and $1$, since we restrict all microscopic analysis to the top Landau level.
Thus, since backscattering is ignored, we can consider the edges of the stripes to behave like the edges of large integer quantum Hall states, with filling factor 1.
In such states, the excitations around an edge of the system are often described by a density operator:
\[
	\rho_{q_x} = \sum_{k_x}c^\dag_{k_x+{q_x}} c_{k_x},
\]
where $c_{k_x}$ annihilates an electron in a one-electron state labelled by $k_x$.
The operator $\rho_{q_x}$ describes a current running in one direction along the edge of the system.  
It obeys the Ka\v{c}-Moody commutation relation ($[\rho_{q_x},\rho_{{q_x}'}] = {q_x} \delta ({q_x}+{q_x}')$) that determines its algebraic behavior and dynamics. 
Thus, on one hand, this operator is defined microscopically, through creation and annihilation operators, and, therefore, can be used in the microscopic Hartree-Fock analysis. 
On the other hand, it can also be described through its algebra as part of a Luttinger liquid description of the system, thus providing a connection between Luttinger liquids and Hartree-Fock.

However, we cannot use this operator to describe excitations localized to a particular edge of the stripe since $\rho_{q_x}$ acts across the entire system. 
When applied to the striped ground state, it would create the same excitations on all edges of all stripes.
To treat the excitations on a particular edge of one of the stripes, we need to localize the operator to that edge.
We do so by introducing an envelope function or ``wavepacket'' function $w_{I \alpha}(k_x)$.  The function $w_{I \alpha}(k_x)$ can have an arbitrary form as long as it equals unity in the vicinity of the edge $\alpha I$ and the different $w_{I \alpha}$ functions have disjoint support  --- meaning only one $w_{I \alpha}(k_x)$ can be nonzero for any given $k_x$  (and $w$ should not have any unintegrable singularities).  
It is important to note that all the physical quantities calculated in this paper are independent of the shape of $w_{I\alpha }(k_x)$ away from the edges, and one never actually needs to worry about the specific form of $w_{I\alpha }$ as a function of $k_x$.

With this ``$w$-wavepacket'', we define a new density operator that affects only the given edge $\alpha I$:
\be
	\rho_{\alpha I,{q_x}} \equiv \sum_{k_x}  w_{I \alpha}(k_x+{q_x}/2) c^\dag_{k_x+{q_x}} c_{k_x}, \label{eq:micro_rho}
\ee
where $w_{\alpha I}(k_x)$ has to be unity within a range of $\pm q_x$ around the edge $\alpha I$ to ensure unitarity of the operator $U^{\varepsilon}_{\alpha I,{q_x}}$ defined below.

The new density operator obeys the Ka\v{c}-Moody commutation relation 
\be
	[\rho_{\alpha I,q_x},\rho _{\beta J,{q_x}'}] = \alpha {q_x} \delta_{IJ} \delta_{\alpha \beta }\delta ({q_x}+{q_x}'),\label{eq:kac-moody}
\ee
just like the ordinary density operator $\rho_{q_x}$, but localized around the given edge. Thus, it is this operator $\rho _{\alpha I,q_x}$ that enters the Luttinger liquid Hamiltonian (Eq.~(\ref{eq:hamilt-mf})) but is also defined in terms of microscopic electron creation operators (Eq.~(\ref{eq:micro_rho})), thus allowing us to make a connection between microscopic Hartree-Fock and the Luttinger liquid Hamiltonian.

To complete our description of low energy excitations around the striped ground state, we construct an operator 
\be 
	U^{\varepsilon}_{\alpha I,{q_x}} \equiv e^{i (\varepsilon \,  \rho_{\alpha I,{q_x}}+\varepsilon^* (\rho_{\alpha I,{q_x}})^\dagger)}
\ee
that creates a state 
\be
	|\alpha I,{q_x};\varepsilon \rangle = U^{\varepsilon}_{\alpha I,{q_x}} |G\rangle  
\ee
when it acts upon the ground state. ($|G\rangle \equiv |G_{\nu^*}\rangle $ further on to simplify notation.)
The operator $U^{\varepsilon}_{\alpha I,{q_x}}$ is constructed by analogy with coherent states of a harmonic oscillator.  
The operator $\rho_{\alpha I,{q_x}}$ acts as a creation operator and $\varepsilon $ as the amplitude of the localized excitation that $\rho_{\alpha I,{q_x}}$ creates.

One can check that $\varepsilon $ does in fact play the role of the amplitude of the localized excitation by calculating the classical expectation value of the electron density operator, $\rho_{J,\beta ,{q_x}'}$, in this state, 
\bea
	\lefteqn{\langle \alpha I,{q_x};\varepsilon |\rho_{\beta J,{q_x}'}|\alpha I,{q_x};\varepsilon \rangle = } \nonumber \\
	& & \alpha \delta_{IJ} \delta_{\alpha \beta } (i\varepsilon {q_x} \delta ({q_x}-{q_x}') - i \varepsilon^* {q_x} \delta ({q_x}+{q_x}')),\nonumber
\eea
for a small $|\varepsilon|$.
This relation implies that the coherent state is indeed a state that corresponds to a classical electron-density excitation on the given edge $\alpha I$.
This density excitation is real and sine-like (cosine-like), with wavevector of magnitude $|q_x|$ if $\varepsilon$ is real (imaginary); the amplitude of the excitation is proportional to $|\varepsilon|$.

Since we substitute what would be a state, in which a complex chiral current runs along the edge $\alpha I$, with two states with real (cosine and sine) density excitations around that edge, we restrict $q_x$ to $q_x>0$ to avoid overcounting. 

One may notice that excitations with $q_x = 0$ cannot be created in this manner. However, this is not a limitation.  
In the absence of backscattering the number of electrons (holes) in each electron (hole) stripe is conserved.
It is easy to see that excitations with $q_x = 0$ violate this constraint and thus cannot be present in our approximation.

Making localized excitations around the edges of different stripes, we can create any possible low energy excitation around the striped ground state.   We describe a set of excitations along the edges ${\alpha I}$ in terms of their $x$-wavevectors $q_{x, \alpha I}$ and their complex amplitudes $\varepsilon_{\alpha I}$.  
The most general form of the coherent state $|\{q_{x, \alpha I}  ; \varepsilon_{\alpha I} \} \rangle $ is one in which there are different density excitations on every edge.  Since the operators $U^\varepsilon_{\alpha I, q_x}$ commute for different $\alpha I$ we can create excitations of multiple edges just by successively applying these operators -- one for each deformed edge.    

In this work, however, we will only be concerned with states in which all edges have the same $x$-wavevector $q_x$ (since it can be easily shown that excitations with different $q_x$ do not interact with each other to quadratic order in the deformations).   We thus define an operator
\bea
U(q_x  &;& \{ \varepsilon_{\alpha I} \})  
=    \prod_{I \alpha} U^{\varepsilon_{\alpha I}}_{\alpha I q_{x}} \\ &=&  
e^{i \sum_{\alpha I} (\varepsilon_{\alpha I} \rho_{\alpha I,{q_{x}}}+\varepsilon^*_{\alpha I} (\rho_{\alpha I,{q_{x}}})^\dag )} 
\eea
which, when applied to the ground state, yields the state
\be
	|q ; \{ \varepsilon_{\alpha I} \} \rangle = U(q_x ; \{ \varepsilon_{\alpha I} \})  
|G\rangle ;
\ee
which has amplitude  $|\varepsilon_{\alpha I} q_{x}|$ on stripe $\alpha I$ and all edges have the same $q_x$.

We will construct and calculate the energy of small-amplitude coherent states in both the Luttinger liquid language and in Hartree-Fock language.
Equating these energies will allow us to determine the full form of $D_{\alpha \beta}({\bf q})$ of the (previously) phenomenological Luttinger liquid Hamiltonian.
Once such a Hamiltonian is written down, it is equivalent to Hartree-Fock in the sense that any of its matrix elements within the subspace defined by  $\rho_{\alpha I,q_x}|G\rangle $ are precisely equal to the corresponding matrix elements calculated in Hartree-Fock.

\subsection{Energy of Coherent States in Luttinger Liquid Language}

The localized density operators $\rho_{\alpha I,{q_x}}$ obey Ka\v{c}-Moody commutation relations and describe currents running along given edges $\alpha I$.  
These fields, therefore, are exactly the fields we need to construct the effective Luttinger liquid Hamiltonian (Eq.~(\ref{eq:hamilt-mf}))
\be
	H = \pi l^2 \sum_{\alpha ,\beta } \sum_{I,J}\int dq_x\alpha \beta \rho_{\alpha I,q_x} D_{\alpha \beta } (-q_x,I-J) \rho_{\beta J,-q_x}\label{eq:locham1}
\ee	
Here the Hamiltonian is Fourier transformed in the $x$-direction, along the stripes.

Semiclassically, the energy of a given state equals the expectation value of the Hamiltonian in this state. 
The excitations of interest to us are the low energy excitations that, in the previous subsection, we represented by the coherent states and whose dynamics the Luttinger liquid Hamiltonian describes.
Given a coherent state $|q_x ; \{ \varepsilon_{\alpha I} \} \rangle$,  the classical expectation value of this energy in this state is then
\be
	E = \langle q_{x} ;  \{ \varepsilon_{\alpha I} \} | H | q_{x} ; \{ \varepsilon_{\alpha I} \} \rangle
\ee

If the interaction matrix, ${\bf D}$, is known, for small values of $|\varepsilon_{\alpha I}|$, we can expand the energy to second order in  $|\varepsilon_{\alpha I}|$ to obtain
\widetext
\Lrule
\bea
	E & = & E_0 + \pi l^2 \sum_{\alpha I,\beta J} \sum_{\gamma M,\delta N} \int dq_x' \gamma \delta D_{\gamma \delta } (-q_x',M-N) \times \nonumber \\
	& & [\langle G | 
		(\varepsilon_{\alpha I} \rho_{\alpha I,{q_{x}}}
		+\varepsilon^*_{\alpha I} (\rho_{\alpha I,{q_{x}}})^\dag )
		\rho_{\gamma M,q_x'} \rho_{\delta N,-q_x'}
		 (\varepsilon_{\beta J} \rho_{\beta J,{q_{x}}}
		+\varepsilon^*_{\beta J}  (\rho_{\beta J,{q_{x}}})^\dag ) 
	|G\rangle \nonumber \\
	& & \mbox -\frac{1}{2}\langle G | 
		(\varepsilon_{\alpha I} \rho_{\alpha I,{q_{x}}}
		+\varepsilon^*_{\alpha I} (\rho_{\alpha I,{q_{x}}})^\dag )
		 (\varepsilon_{\beta J} \rho_{\beta J,{q_{x}}}
		+\varepsilon^*_{\beta J}  (\rho_{\beta J,{q_{x}}})^\dag ) 
		\rho_{\gamma M,q_x'} \rho_{\delta N,-q_x'}
	|G\rangle \nonumber \\
	& & \mbox -\frac{1}{2}\langle G | 
		\rho_{\gamma M,q_x'} \rho_{\delta N,-q_x'}
		(\varepsilon_{\alpha I} \rho_{\alpha I,{q_{x}}}
		+\varepsilon^*_{\alpha I} (\rho_{\alpha I,{q_{x}}})^\dag )
		 (\varepsilon_{\beta J} \rho_{\beta J,{q_{x}}}
		+\varepsilon^*_{\beta J}  (\rho_{\beta J,{q_{x}}})^\dag ) 
		|G\rangle ],\label{eq:energy-ll}
\eea 
\Rrule
\narrowtext
\noindent where the constant $E_0 = \langle G | H |G \rangle$ can be considered to be the ground state energy. This ground state energy is a large additive constant that is the same for all excitations.  
It appears because the coherent state describes the entire excited ``striped'' state and not just the Luttinger liquids on the edges of the stripes (the set of Luttinger liquids can be thought of as the excited ``striped'' state minus the ``striped'' ground state).   Given values of the coefficients $\varepsilon_{I \alpha}$, 
Eq. (\ref{eq:energy-ll}) can be  evaluated using the commutation relation of the density operators $\rho_{\alpha I,q} $ given by Eq.~(\ref{eq:kac-moody}).

We now create an excitation of wavevector ${\bf q}=(q_x,q_y)$ by taking $\varepsilon_{\alpha I}$ to be given by 
\be
	\varepsilon_{I\alpha}=\eta e^{i\alpha \frac{\theta_S}{2}}e^{iq_ya(I+\alpha\frac{\nu^*}{2})} \label{eq:waveofexcitations}
\ee
where $\eta$ is the overall amplitude of the excitation.    We recall that the complex phase of the parameters $\varepsilon$ determine the phase of the excitation along the edge --- representing a sine excitation if $\varepsilon$ is real or a cosine if $\varepsilon$ is imaginary.  Thus, when the 
constant $\theta_S = 0$ we have a  symmetric (smectic) excitation; when $\theta_S=\pi $ we have an antisymmetric excitation; and for $\theta_S =  \pi /2$ we have a ``mixed'' excitation, in which there are a cosine excitations on one side of all stripes and sine excitations on the opposite side.   Inserting this form into Eq. (\ref{eq:energy-ll}) yields
\bea
	E & = & E_0 + \frac{(\eta q_x l)^2 }{2} \frac{L_x N}{a} \times \nonumber \\
	& & \left[ D_{LL}({\bf q}) + D_{RR}({\bf q}) + 2 \mbox{Re}\{ e^{i\theta_S+iq_ya\nu^*} D_{RL}({\bf q})\} \right] \nonumber.
\eea
yielding the following three energies for  symmetric, antisymmetric, and ``mixed'' excitations
\bea
	\Delta E_S & = & \frac{(\eta q_x l)^2 }{2} \frac{L_x N}{a} \sum_{\alpha \beta} e^{iq_ya\frac{\nu^*}{2}(\alpha -\beta )}D_{\alpha \beta } ({\bf q}) \nonumber \\
	\Delta E_A & = & \frac{(\eta q_x l)^2 }{2} \frac{L_x N}{a} \sum_{\alpha \beta} \alpha \beta e^{iq_ya\frac{\nu^*}{2}(\alpha -\beta )}D_{\alpha \beta } ({\bf q}) \label{eq:threeen}\\ \nonumber	
	\Delta E_M & = & \frac{(\eta q_x l)^2 }{2} \frac{L_x N}{a} [D_{LL} ({\bf q}) + D_{RR} ({\bf q}) 
	\\ & & \mbox{\hspace*{60pt}} - 2 \mbox{Im} \{e^{iq_ya\nu^*} D_{RL} ({\bf q}) \} ] . \nonumber
\eea

If the interaction matrix ${\bf D}$ is known, we can obtain the energies of any small excitation using the Luttinger liquid description.
Conversely, it is clear that, if the energies $E$ are known for arbitrary (small) $\varepsilon $, the matrix ${\bf D}$ may be obtained.
We shall calculate these energies in the Hartree-Fock approximation for the coherent states that we can describe microscopically as well as through the density operators (Sec.~\ref{sec:parameters}).

\section{Energy of Coherent States}
\label{sec:parameters}

\subsection{Coherent States in Hartree-Fock Language}

To find the energies $E_S$, $E_A$ and $E_M$ (Eq.~(\ref{eq:threeen})) from first principles, we need to describe the low energy excited states within Hartree-Fock.
This also can be done using the coherent states, since the formalism can be easily translated into the Hartree-Fock language. 
To determine the energies, it will suffice to consider the states where $q_x$ is the same for every edge. 
Within the Hartree-Fock approximation, the operator $
U(q_x  ; \{ \varepsilon_{\alpha I} \}) $ 
is a rotation in the 
Hilbert space of the wavefunctions $\{ \psi_{L,{k_x}_n}\}$ of the form
\be
	U(q_x  ; \{ \varepsilon_{\alpha I} \})   =  e^{i R(q_{x}  ; \{ \varepsilon_{\alpha I} \})} 
\ee
where $R(q_{x}  ; \{ \varepsilon_{\alpha I} \})$ is a matrix whose elements are given by
\bea
	\lefteqn{R_{k_x,k_x'}(q_{x}  ; \{ \varepsilon_{\alpha I} \}) = } \nonumber \\
	& = & \delta_{k_x',k_x-q_x} W(k_x-\frac{q_x}{2}) +  \delta_{k_x',k_x+q_x} W^*(k_x+\frac{q_x}{2}),
\eea
where 
\be
W(k_x) = \sum_{\alpha I} \varepsilon_{\alpha I} w_{\alpha I}(k_x)  
\ee
and $w_{\alpha I}(k_x)$ is the wavepacket envelope function defining the operator $\rho_{I \alpha}$ in Eq. (\ref{eq:micro_rho}).  

\begin{figure*}
	\centering
	\includegraphics{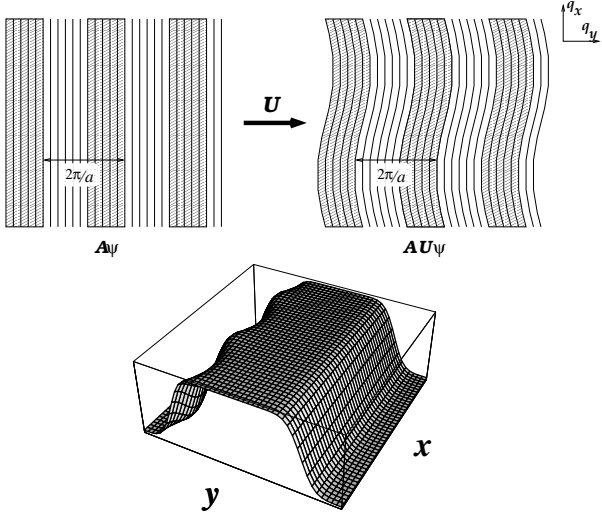}
	\caption{\label{fig:rotation} Top figure is a schematic representation of the  $U(q_x  ; \{ \varepsilon_{\alpha I} \}) $ transformation.   Each single-electron wavefunction, represented by a solid line here, is transformed by  $U(q_x  ; \{ \varepsilon_{\alpha I} \}) $.  Then a Slater determinant, ${\cal A}_{a,\nu^*}\tilde{\Psi}_{q_x}^{\varepsilon_{\alpha I}}$, is created, in which the new single-electron wavefunctions are filled in a striped fashion, just like in the unperturbed case.  It is important to note here that, using the transformation $U(q_x  ; \{ \varepsilon_{\alpha I} \}) $, we can also create states that have different excitations on different edges of the system.  
The electron-density profile in the bottom figure illustrates this point.  Here, an excitation is created on the left edge of one stripe (we used the L=0 single-electron functions for simplicity).}
\end{figure*}

The rotation $U(q_x  ; \{ \varepsilon_{\alpha I} \}) $ is a unitary operator that preserves orthogonality and completeness of $\{ \psi_{L,{k_x}_n}\} $, as long as the $w$-wavepacket is unity over a range of $\pm q_x$ around the edge.  The Slater determinant  ${\cal A}_{a,\nu^*}\tilde{\Psi}_{q_x}^{\varepsilon_{\alpha I}}$, in which $\tilde{\Psi}_{q_x}^{\varepsilon_{\alpha I}} = U(q_x  ;  \{ \varepsilon_{\alpha I} \}) \Psi_L$ and all the states are filled in a striped fashion, represents the excited coherent state $ | q_x ; \{\varepsilon_{\alpha I} \} \rangle$ in the Hartree-Fock approximation (Fig.~\ref{fig:rotation}).
Later in this Section, we will calculate the energy of the
the Slater determinant ${\cal A}_{a,\nu^*}\tilde{\Psi}_{q_x}^{\varepsilon_{\alpha I}}$ for $\varepsilon_{\alpha I}$ corresponding to symmetric, antisymmetric, and mixed excitations (See Eq. \ref{eq:waveofexcitations}).  In this way we will find the energies $\Delta E_S$, $\Delta E_A$, and $\Delta E_M$ and read off the full interaction matrix of the Luttinger liquid theory, ${\bf D}({\bf q})$.

\subsection{Ground-State Energy}

In the space of the one-electron wave-functions $\{ \psi_{{k_x}_n}\} $, the Coulomb potential screened by the lower, completely filled Landau levels, $V_L({\bf r}),$\cite{aleiner1} takes on a matrix form
\bea
	V_{mnst} & = & \int_{{\bf r},{\bf r'}} \psi_{{k_x}_m}({\bf r}) \psi^*_{{k_x}_n}({\bf r})V_L({\bf r-r'}) \psi_{{k_x}_s}({\bf r'}) \psi^*_{{k_x}_t}({\bf r'}) \nonumber \\
 	\mbox{} & = & \frac{2\pi }{L_x^2}\delta ({k_x}_m-{k_x}_n+{k_x}_s-{k_x}_t) \times \nonumber \\
	& \int & \frac{dp}{2\pi } e^{-ipl^2 \frac{{k_x}_m+{k_x}_n-{k_x}_s-{k_x}_t}{2} } V_{\mbox{eff}}({k_x}_n-{k_x}_m,p) \label{eq:delta-fn},
\eea
where
\be
	V_{\mbox{eff}}(q_x,q_y) = V_L(q) \left( L_L\left( \frac{q^2l^2}{2}\right) \right)^2 e^{-\frac{q^2l^2}{2}}; 
\ee
$q = \sqrt{q_x^2+q_y^2}$ and $L_L(q)$ is the $L$th Laguerre polynomial.
The screened Coulomb potential, $V_L({\bf r})$, is related to the bare Coulomb potential, $V({\bf r})=2\pi e^2/(\kappa q)$ with $\kappa$ the background dielectric constant as
\be
	V_L(q)=\frac{2\pi e^2}{q}\frac{1}{\kappa \epsilon (q)},
\ee
with the correction term\cite{aleiner1},
\be
	\epsilon (q)=1+ \frac{2}{q a_B}[1-J_0^2(qR_c)], \label{eq:dielectric}
\ee
arising from screening of Coulomb interaction between electrons in the partially filled upper Landau level by the electrons in the completely filled lower Landau levels.
Here $a_B=\hbar ^2 \kappa /m^*e^2$ is the Bohr radius and $m^*$ is the effective electron mass in the system. 
Throughout the paper, $J_\mu (x)$ is the Bessel function of order $\mu $. 
The form of the dielectric constant in Eq.~(\ref{eq:dielectric}) is obtained through a semiclassical approximation and is valid for $L >> 1$.  
A fuller expression for the effective dielectric constant that is valid for small $L$ is given in Ref.~\onlinecite{aleiner1}.

The Hartree-Fock energy of the ground state is therefore
\be
	E_{GS} = \frac{1}{2}\frac{L_x^2}{(2\pi )^2}\sum_{I,J} \int^{k_{RI}}_{k_{LI}} d{k_x}_m \int^{k_{RJ}}_{k_{LJ}} d{k_x}_n \left( V_{mmnn}-V_{mnnm} \right) \label{eq:gsenergy}, 
\ee
where $k_{\alpha I} \equiv a/l^2(I+\alpha \nu^*)$. 
The first term of Eq.~(\ref{eq:gsenergy}) is the direct component of the ground state energy, $E^D_{GS}$, and the second term is the exchange component, $E^X_{GS}$.
The superscripts $D$ and $X$ will label direct and exchange energies henceforth.
After substitution of Eq.~(\ref{eq:delta-fn}) into Eq.~(\ref{eq:gsenergy}) and some mathematical manipulations, Eq.~(\ref{eq:gsenergy}) becomes\cite{macdonald2}
\bea
	E_{GS} & = & \frac{1}{2}\frac{L_xNa}{2\pi l^2} \sum_n U(\frac{2\pi n}{a}) \left( \frac{\sin (\pi \nu^* n)}{\pi n} \right)^2 \label{eq:gs}\\
	U(q) & = & \frac{V_{\mbox{eff}}(0,q)}{2\pi l^2} - \int \frac{du\, dp}{(2\pi )^2} e^{iql^2u} V_{\mbox{eff}}(u,p)\nonumber,
\eea
where the ratio $L_x Na/2\pi l^2$ is the number of states in the partially filled Landau level and $V_{\mbox{eff}}(0,0)$ is assumed to be 0, to account for the uniform positive background.

The width of the stripes is fixed by minimization of the ground state energy (Eq.~(\ref{eq:gs})), i.e. by the requirement that $\partial_a E_{GS} = 0$.

\subsection{Hartree-Fock Energy of Excitations}

Following the procedure given in Sec.~\ref{sec:excitations}, we represent the excitations by coherent states.
Within the coherent state formalism, any low energy excited state turns out to be conveniently represented by a single Slater determinant, ${\cal A}_{a,\nu^*}\tilde{\Psi}_{q_x}^{\varepsilon_{\alpha I}}$, where $\tilde{\Psi}_{q_x}^{\varepsilon_{\alpha I}} = U(q_x  ; \{\varepsilon_{\alpha I} \}) \Psi$ (the Hilbert-space vector $\Psi$ is defined by Eq.~(\ref{eq:psi})), that can be used as the Hartree-Fock wavefunction.
The transformed single-electron states are
\bea
	\tilde{\psi}_{k_x} & \approx & 
		\psi_{k_x} \left( 1-|W(k_x+\frac{q_x}{2})|^2 - |W(k_x-\frac{q_x}{2})|^2\right) \label{eq:psi_tilde}\\
		& & \mbox{} + i W(k_x+\frac{q_x}{2})\psi_{k_x+q_x}+i W^*(k_x-\frac{q_x}{2}) \psi_{k_x-q_x}.\nonumber \\
		& &  \mbox{} - W(k_x+\frac{q_x}{2})W(k_x+\frac{3 q_x}{2}) \psi_{k_x+2 q_x} \nonumber \\
		& & \mbox{}- W^*(k_x-\frac{q_x}{2}) W^*(k_x-\frac{3q_x}{2}) \psi_{k_x-2q_x}\nonumber 
\eea
to second order in the small amplitude ($W$ or $\epsilon_{\alpha I}$).   To the same order, the matrix element of the renormalized 
\vspace{3pt}

\Lrule

\vspace{6pt}

\noindent
Coulomb potential in this set of states is
\bea
	\tilde{V}_{mnst} & = & \int_{{\bf r},{\bf r'}}\, V_L({\bf r-r'})\tilde{\psi }_{{k_x}_m}({\bf r}) \tilde{\psi }^*_{{k_x}_n}({\bf r}) \tilde{\psi }_{{k_x}_s}({\bf r'}) \tilde{\psi }^*_{{k_x}_t}({\bf r'}) \nonumber \\
 	\mbox{} & = & V_{mnst} + \sum_{b=\pm 1} [ \nonumber \\
	& + & W({k_x}_m+b\frac{q_x}{2}) W^*({k_x}_n+b\frac{q_x}{2}) V_{(m+bq_x)(n+bq_x)st} + \nonumber \\
	& + & W ({k_x}_s+b\frac{q_x}{2}) W^*({k_x}_t+b\frac{q_x}{2}) V_{mn(s+bq_x)(t+bq_x)} + \nonumber \\
	& + & W({k_x}_m+b\frac{q_x}{2}) W^*({k_x}_s-b\frac{q_x}{2}) V_{(m+bq_x)n(s-bq_x)t} + \nonumber \\
	& + & W({k_x}_n-b\frac{q_x}{2}) W^*({k_x}_t+b\frac{q_x}{2}) V_{m(n-bq_x)s(t+bq_x)}]. \nonumber
\eea
It is easy to see that, because of the $\delta $-function in Eq.~(\ref{eq:delta-fn}), the last two terms of the expansion of $\tilde{\psi}_{k_x}$ (Eq.~(\ref{eq:psi_tilde})) do not contribute to the matrix element $\tilde{V}_{mnst}$ until the fourth order in $W$.  Effectively, we need to expand $\tilde{\psi}_{k_x}$ only to first order in $W$ and normalize it.

The Hartree-Fock energy of given excited state is calculated the same way as the ground state energy (Eq.~(\ref{eq:gsenergy})) but with $(\tilde{V}_{mmnn}-\tilde{V}_{mnnm})$ substituted for $(V_{mmnn}-V_{mnnm})$.
Slightly tedious but straightforward computation leads to the Hartree-Fock energies for the symmetric ($\Delta E_S$), antisymmetric ($\Delta E_A$), and mixed ($\Delta E_M$) excitations above ground state energy. 
If we define two functions
\widetext
\bea
	F(k_x,k_y) & = &\frac{V_{\mbox{eff}}(k_x,k_y)}{2\pi l^2}-\int \frac{dp\, du}{(2\pi )^2} \cos (p l^2 k_x)  \cos (u l^2 k_y)  V_{\mbox{eff}}(u,p), \\
	\Delta_s (k_x,k_y) & = & \frac{4 \sin^2(\frac{k_x l^2 k_y}{2})}{(k_x l^2 k_y)^2},
\eea
we get a greatly simplified expression for the energies
\bea
	\Delta E_S & = & 2 |\eta |^2 \frac{L_xN}{2\pi a}(q_x l)^2 \sum_n \sin^2(\pi n\nu^*)(\Delta_s (q_x,\frac{2\pi n}{a}+q_y)F(q_x,\frac{2\pi n}{a}+q_y)-\Delta_s (q_x,\frac{2\pi n}{a})F(0,\frac{2\pi n}{a})), \label{eq:symm}\\
	\Delta E_A & = & 2 |\eta |^2 \frac{L_xNa}{2\pi a}(q_x l)^2 \sum_n ( \cos^2(\pi n\nu^*)\Delta_s (q_x,\frac{2\pi n}{a}+q_y)F(q_x,\frac{2\pi n}{a}+q_y)- \sin^2(\pi n\nu^*)\Delta_s (q_x,\frac{2\pi n}{a})F(0,\frac{2\pi n}{a})), \label{eq:antisymm}\\
	\Delta E_{M_0} & = & |\eta |^2 \frac{L_xN}{2\pi a}(q_x l)^2 \sum_n \sin (2 \pi n\nu^*) \Delta_s (q_x,\frac{2\pi n}{a}+q_y)F(q_x,\frac{2\pi n}{a}+q_y),  \label{eq:mixed}\\
	\Delta E_M & = & \frac{\Delta E_S+\Delta E_A}{2}+\Delta E_{M_0} \nonumber.
\eea
Here we used exactly the same coherent states to represent the three basic excitations, as we did in Eq.~(\ref{eq:threeen}) to obtain the general form of $\Delta E_S$, $\Delta E_A$, and $\Delta E_{M_0}$ from the Luttinger liquid Hamiltonian (note that it is $\Delta E_{M_0}$ and not $\Delta E_{M}$ that is proportional to $\mbox{Im}\{ e^{iq_ya\nu^*} D_{RL}({\bf q}) \}$).
Using Eqs.~(\ref{eq:threeen}) and (\ref{eq:sym-ant}), we can now simply read off the elements of the interaction matrix ${\bf D}({\bf q})$ from Eqs.~(\ref{eq:symm})--(\ref{eq:mixed}):
\bea
	D_S ({\bf q}) & = & \frac{1}{2\pi } \sum_n \left[ \sin^2(\pi n\nu^*)\Delta_s (q_x,\frac{2\pi n}{a}+q_y)F(q_x,\frac{2\pi n}{a}+q_y)-\sin^2(\pi n\nu^*)\Delta_s (q_x,\frac{2\pi n}{a})F(0,\frac{2\pi n}{a})\right] , \label{eq:ds_res}\\
	D_A({\bf q}) & = & \frac{1}{2\pi } \sum_n \left[ \cos^2(\pi n\nu^*)\Delta_s (q_x,\frac{2\pi n}{a}+q_y)F(q_x,\frac{2\pi n}{a}+q_y)- \sin^2(\pi n\nu^*)\Delta_s (q_x,\frac{2\pi n}{a})F(0,\frac{2\pi n}{a})\right] , \label{eq:da_res}\\
	\mbox{Im}\{ e^{iq_ya\nu^*} D_{RL}({\bf q}) \} & = & -\frac{1}{2\pi } \sum_n \sin (2 \pi n\nu^*) \Delta_s (q_x,\frac{2\pi n}{a}+q_y) F(q_x,\frac{2\pi n}{a}+q_y),  \label{eq:dm_res}
\eea
\Rrule
\narrowtext
These expressions for the Luttinger liquid interaction matrix ${\bf D}({\bf q})$ are valid for any wavevector $q_y$ and for small $q_x$. 

Now, we can expand Eqs.~(\ref{eq:ds_res})--(\ref{eq:dm_res}) for $q_x l$ and $q_y l << 1$ and re-write them in terms of the hydrodynamic parameters $Y$, $K$, $\chi $, and $C$, thus completing the connection between Luttinger liquids and hydrodynamics and obtaining all the hydrodynamic parameters. 
\Lrule

\vspace{12pt}

\section{Parameters of the Hydrodynamic Theory}
\label{sec:results}

Expansion of Eq.~(\ref{eq:symm}) gives the expected form for the symmetric component of the effective interaction
\be
	D_S({\bf q}) = \frac{l^2 a^2}{4} (Y q_y^2 + K q_x^4), \label{eq:dsym}
\ee
where the coefficients $Y$ and $K$ are
\widetext
\bea
	Y & = & \frac{1}{\pi a^2}\sum_n \sin^2(\pi \nu^* n)
	\left[ 
		\frac{1}{2\pi l^4} \left. \partial^2_qV_{\mbox{eff}}(0,q) \right|_\frac{2\pi n}{a} + \frac{l^2}{2} \int \frac{d\xi }{2\pi }V_{\mbox{eff}}(0,\xi )\xi^3\left( J_0(\xi \frac{2\pi n}{a} l^2)-J_2(\xi \frac{2\pi n}{a} l^2)\right) 
	\right], \label{eq:y} \\
	K & = & \frac{1}{16\pi } \sum_n \frac{\sin^2(\pi \nu^* n)}{(\pi n)^2} 
	\left[ 
		\frac{1}{2\pi l^4} \left( 
			\left. \partial^2_qV_{\mbox{eff}}(0,q)\right|_\frac{2\pi n}{a} 
			- \frac{a}{2\pi n} \left. \partial_qV_{\mbox{eff}}(0,q)\right|_\frac{2\pi n}{a} 
			\left( 1+\frac{1}{3} 
				\left( \frac{2\pi n l}{a} \right)^4 \right)\right)
		\right.	- \nonumber \\
	& & - \frac{(2\pi n)^3l^4}{3a^3}\int \frac{d\xi }{2\pi }V_{\mbox{eff}}(0,\xi )\xi^2 J_1(\xi \frac{2\pi n}{a} l^2) - \nonumber \\
	& & - \left. \frac{(2\pi n) l^4}{4a}\int \frac{d\xi }{2\pi }V_{\mbox{eff}}(0,\xi )\xi^4 \left( J_1(\xi \frac{2\pi n}{a} l^2)+J_3(\xi \frac{2\pi n}{a} l^2) \right)
	\right] \label{eq:k}.
\eea
\Rrule
\narrowtext
Note there there is no $q_x^2$ term in expansion of the energy of the symmetric excitation as follows from symmetry arguments for a smectic excitation (Fig.~\ref{fig:shear}).  For interactions with circular  symmetry, the $q_x^2$-term is proportional to  $\partial_aE_{GS}$ that vanishes when we minimize the ground state energy with respect to the stripe-width, $a$.

The effective antisymmetric component of the interaction (from Eq.~(\ref{eq:antisymm})) behaves as
\be
	D_A({\bf q}) = \frac{1}{2\pi }\frac{V (q)+\chi^{-1}}{2\pi l^2}, \label{eq:da}
\ee
in accord with the phenomenological arguments given in Sec.~\ref{sec:lowenergy}. 
The constant $\chi^{-1}$ turns out to be
\bea
	\chi^{-1} & = & \sum_n \cos (2\pi n\nu^* )
	\left[ 
		V_{\mbox{eff}}(0,\frac{2\pi n}{a})
	\right. \label{eq:chi} \\
	& & 
	\left.
		- 2\pi l^2 \int \frac{d\xi }{2\pi }V_{\mbox{eff}}(0,\xi )\xi J_0(\xi \frac{2\pi n}{a} l^2)
	\right], \nonumber \\
	& = & 2\pi l^2 \left. \frac{\partial^2 E_{GS}}{\partial \nu^2 }\right|_a \nonumber
\eea
as expected from Eqs.~(\ref{eq:chiinv}) and ~(\ref{eq:gs}).
If the 2DEG is screened by a metal plate placed a distance $d$ away from the surface of the gas, then $V(q)$ in Eq.~(\ref{eq:da}) is replaced by $V(0)=\lim_{q \to 0} V(q) $.  
The screened Coulomb potential has the form $V(q) = \frac{2\pi e^2}{\kappa q}(1-e^{-dq})$, hence $V(0) = 2\pi e^2 d/\kappa$.

The off-diagonal term $\mbox{Im} \{ e^{iq_ya\nu^*} D_{RL} ({\bf q}) \}$, given by the expansion of Eq.~(\ref{eq:mixed}), is linear in $q_y$ when the wavevector is small:
\be
	\mbox{Im}\{ e^{iq_ya\nu^*} D_{RL}({\bf q}) \} = \frac{ma}{2\pi }Cq_y, \label{eq:doffdiag}
\ee
where the constant $C$ is
\bea
	C & = & -\frac{l}{ma} \sum_n \sin (2\pi n \nu^*) 
	\left[ 
		\left. \frac{1}{2\pi l^3}\partial_qV_{\mbox{eff}}(0,q)\right|_\frac{2\pi n}{a} 
	\right.+ \nonumber \\
	& & 
	\left.
		+l\int \frac{d\xi }{2\pi }V_{\mbox{eff}}(0,\xi )\xi^2 J_1(\xi \frac{2\pi n}{a} l^2)
	\right].	\label{eq:c}
\eea
Note that the constant $C$ and the compression elastic modulus $Y$ satisfy the relation  $C=-\frac{Y}{ma}\frac{\partial a}{\partial n_0}$, as, again, was expected from phenomenological considerations above (See Eq. (\ref{eq:C})).

\section{Dynamics}
\label{sec:dynamics}

As the first, simplest application of the model, we can consider the dynamical properties of the striped state.

Using the commutation relations of the density operators, Eq.~(\ref{eq:kac-moody}), we can find the Heisenberg equations of motion for the density operators
\bea
	i \partial_t \rho_{\alpha {\bf q}} & = & [\rho_{\alpha {\bf q}}, H] \\
	i \partial_t \rho_{\alpha {\bf q}} & = & \frac{2\pi l^2}{a} q_x \sum_{\beta }\beta D_{\alpha \beta }({\bf q}) \rho_{\beta {\bf q}}
\eea
Fourier-transformation and diagonalization of the equation of motion gives us the dispersion relation (keeping in mind that, by symmetry, $D_{RR}({\bf q})=D_{LL}({\bf q}))$:	 
\be
	\omega = \pm \frac{2\pi l^2}{a} q_x \sqrt{D_{RR}({\bf q})D_{LL}({\bf q})-D_{RL}({\bf q})D_{LR}({\bf q})}.
\ee
The expression under the square root is the determinant of the interaction matrix, $\det {\bf D(q)}$.  Since the determinant of a matrix does not change with the basis in which the matrix is defined, for any basis of density operators that we choose for our Hamiltonian, the dispersion relation is
\be
	\omega = \pm \frac{2\pi l^2}{a} q_x \sqrt{\det {\bf D(q)}}.
\ee
For example, it is convenient for us to use $\rho_{A {\bf q}}$ and $\rho_{S {\bf q}}$ for further calculations.  
In this basis, the dispersion is
\be
	\omega = \pm \frac{4\pi l^2}{a} q_x \sqrt{D_S({\bf q})D_A({\bf q})-\frac{1}{4}(\mbox{Im} \{e^{iq_ya\nu^*} D_{RL} ({\bf q})\})^2},
\ee
and we can use the parameters and equations found in the previous section to study the behavior of the dispersion more quantitatively.

\begin{figure*}[t]
	\centering
	\scalebox{0.5}{\includegraphics{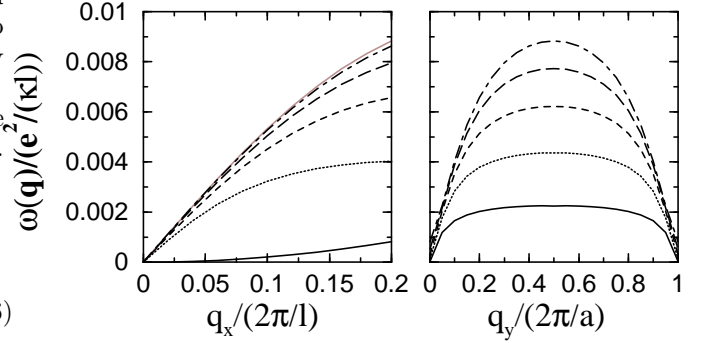}}
	\caption{\label{fig:parameters}
Dispersion relation of a striped state with $\nu = 4.45$ ($L = 2$, $\nu^* = 0.45$, {\it i.e.} a filling fraction around 9/2).  
Coulomb interactions here are assumed to be screened only by the lower, fully filled, Landau levels.  
This screening depends on the density of electrons in the system, which is taken to be $n_s = 2.67\times 10^{11} \mbox{cm}^{-2}$. 
In the left plot, the dispersion relation is given as a function of $q_x$ for fixed values of $q_y$: $q_y = 0$ is the solid line, $0.1(2\pi /a)$ is the dotted line,  $0.2(2\pi /a)$ is the dashed line, $0.3(2\pi /a)$ is the long-dashed line, $0.4(2\pi /a)$ is the dot-dashed line, and $0.5(2\pi /a)$ is the grey line.  Here, the distance between the centers of the stripes that minimizes the energy of the striped ground state is $a = 6.54 l $. 
In the right plot, the dispersion relation is given as a function of $q_y$ for fixed values of $q_x$: $0.04(2\pi/l)$ is the solid line, $0.08(2\pi/l)$is the dotted line, $0.12(2\pi/l)$ is the dashed line, $0.16(2\pi/l)$ is the long-dashed line, and $0.2(2\pi/l)$ is the dot-dashed line.  Note that for $q_x \neq 0$, the curves do not reach $\omega = 0$ for $q_y = 0$ and $2\pi/a$ but come down to a very small value  $\omega = q_x^3 l^2 \sqrt{K(V(q_x)+\chi^{-1})}$ (that corresponds to the solid line in the left plot).}
\end{figure*}

In Figure \ref{fig:parameters}, we plot the positive-$q_x$ branch of the dispersion as a function of $(q_x,q_y)$, for small values of $q_x$ and for $q_y$ within the Brillouin zone $(0,2\pi /a)$;  the other parameters of the system are specified in the caption.                      .
Note that, even though the functions $D_S({\bf q})$, $D_A({\bf q})$, and $\mbox{Im} \{e^{iq_ya\nu^*} D_{RL} ({\bf q})\}$ are not periodic in $q_y$, it is easy to show that the determinant, $\det {\bf D(q)}$, is periodic, in any basis of choice.

The modes corresponding to the positive and negative branches of the dispersion are $\rho_{+,{\bf q}}$ and $\rho_{-,{\bf q}}$, respectively.  
They can be found to be
\bea
	\rho_{+,{\bf q}} & = & \frac{1}{N} (D_{A}({\bf q}) \rho_{S{\bf q}} \nonumber \\
			& & \mbox{}-(\frac{i}{2}\mbox{Im} \{e^{iq_ya\nu^*} D_{RL} ({\bf q})\}-\sqrt{\det {\bf D(q)}})\rho_{A{\bf q}}) \\
	\rho_{-,{\bf q}} & = & \frac{1}{N} (D_{A}({\bf q}) \rho_{S{\bf q}} \nonumber \\
			& & \mbox{}-(\frac{i}{2}\mbox{Im} \{e^{iq_ya\nu^*} D_{RL} ({\bf q})\}+\sqrt{\det {\bf D(q)}})\rho_{A{\bf q}}) \\
	N^2 & = & D_{A}({\bf q})(D_{A}({\bf q})+D_{S}({\bf q}))
\eea

\section{Tunneling Properties}
\label{sec:tunneling}

The Luttinger liquid model also gives some interesting results about the properties of perpendicular tunneling into the striped state.
Following the usual treatment of electrons in one dimension, we represent the operator for an electron ``running along'' the edge $\alpha I$, $\Psi_{\alpha I}$, through bosonic fields $\phi_{\alpha I}$, so that $\Psi_{\alpha I} \sim e^{i \phi_{\alpha I}}$.
The bosonic fields are related to the density operators as $\rho_{\alpha I}(x)=\alpha \partial_x \phi_{\alpha I}(x)/2\pi $.
These identifications allow us to calculate the electron Green's function
\be
	G_e(\tau ) = \langle \Psi_{\alpha I}(x,0) \Psi^\dagger_{\alpha I}(x,\tau ) \rangle \propto  e^{-{\cal C}(\tau )},
\ee
where
\be
	{\cal C}(\tau ) = -(\langle \phi_{\alpha I}(x,0) \phi_{\alpha I}(x,\tau ) \rangle -\langle \phi_{\alpha I}(x,0) \phi_{\alpha I}(x,0) \rangle ).
\ee

To get the time-dependent correlations, we follow MF and construct an imaginary-time action from the Luttinger liquid Hamiltonian (Eq.~(\ref{eq:hamilt-mf}))\cite{macdonald1}.
In terms of the bosonic fields $\phi_{\alpha I}$, the imaginary-time action takes the form
\be
	S_0 = \frac{1}{2}\int \frac{d{\bf q} d\omega }{(2\pi )^3}\phi_{\alpha }({\bf q},\omega ) M_{\alpha \beta }(-{\bf q},-\omega ) \phi_{\beta }(-{\bf q},-\omega ) \label{eq:unperturbed},
\ee
where
\be
	M_{\alpha \beta }({\bf q},\omega ) = \alpha \delta_{\alpha \beta } \frac{i\omega q_x}{2\pi a} +\frac{(q_xl)^2}{a^2}D_{\alpha \beta }({\bf q}). \label{eq:m}
\ee
The minor differences between Eq.~(7) of Ref.~\onlinecite{macdonald1} and  Eq.~(\ref{eq:m}) here are due to the differences between our Fourier-transform conventions.

The bosonic Green's function is just the inverse of the kernel $M_{\alpha \beta }({\bf q},\omega )$:
\bea
	\lefteqn{\langle \phi_{\alpha }({\bf q},\omega ) \phi_{\beta }(-{\bf q},-\omega ) \rangle = {\bf M}^{-1}({\bf q},\omega ) = } \nonumber \\
	& = & \frac{-\alpha \delta_{\alpha \beta }\frac{i\omega q_x}{2\pi a}+\frac{(q_xl)^2}{a^2}D_{\alpha \beta }({\bf q}) }{\frac{\omega^2q_x^2}{4\pi^2a^2}+\frac{(q_xl)^4}{a^4}\det {\bf D}({\bf q})} \label{eq:green1}.
\eea
Fourier transformation of Eq.~(\ref{eq:green1}) gives an expression for the real-space bosonic Green's function that, after integration over $\omega $ takes the form
\bea
	\lefteqn{\langle \phi_{\alpha I}(x,0) \phi_{\alpha I}(x,\tau ) \rangle =} \nonumber \\
	& = & -\alpha \, a\frac{\mbox{sign}\tau }{4\pi } \int \frac{d{\bf q}}{q_x}e^{-|q_x|\frac{2\pi l^2}{a}\sqrt{\det {\bf D}}\tau } \\
	& & + \frac{a}{2} \int \frac{d{\bf q}}{|q_x|} \frac{D_{\alpha \alpha}({\bf q})}{\sqrt{\det {\bf D} ({\bf q})}} e^{-|q_x|\frac{2\pi l^2}{a}\sqrt{\det {\bf D}}\tau } \label{eq:omegaintegration}.
\eea
The first integral vanishes if one assumes that it should be taken as a principal value.
The second integral can be evaluated approximately for large values of $\tau $, at which our low-energy theory is valid, to give us the low energy contribution to the bosonic Green's function and, thus, to the electron Green's function, $G_e (\tau )$.
It is important to note, however, that we should keep to energies large enough (times low enough) to be able to ignore possible backscattering effects.  The energies that we consider must be higher than the backscattering energy, $1/t_b$, where $t_b$ is the backscattering time.

Note that if the excitations along the edges of the stripes were independent of each other, $\det {\bf D}({\bf q})$ would equal $D^2_{\alpha \alpha }$ and we would get exactly the logarithmic form expected of ${\cal C}(\tau )$ for a system of non-interacting Luttinger liquids.
Specifically, we would find $G_e(t) \propto 1/t$, as required from a set of non-interacting $\nu = 1$ edges.
However, the excitations on the edges of the striped state do interact with each other, and the tunneling exponent behaves differently from a noninteracting Luttinger liquid.

For unscreened Coulomb interactions, the contribution of the low energy excitations to ${\cal C}(\tau )$ (Eq.~(\ref{eq:omegaintegration})) can be obtained using the approximate low wavevector form of ${\bf D}({\bf q})$ given by Eqs.~(\ref{eq:dsym}), (\ref{eq:da}), and (\ref{eq:doffdiag}),
\be
	{\cal C} (\tau ) = \frac{1}{2\pi l^2}\sqrt{\frac{V_Cl}{Y}}\int \frac{d{\bf q}}{|q_xq_y| \sqrt{q}} (1 - e^{-\frac{|q_xq_y|}{\sqrt{q}}\frac{l^2}{2}\sqrt{YV_Cl}\tau }),\label{eq:unscrgreen}
\ee
where $V_C = 2\pi l^2/(l\kappa )$ is the characteristic Coulomb energy of electrons separated by the magnetic length $l$, and $Y$ is given in Eq.~(\ref{eq:y}).
To get the approximate form of the interaction matrix, we assumed that $q$ is low enough that $V(q) >> \chi^{-1}$ and $V(q)Y >> (mC)^2/4 $ (the expression for $C$ is given in Eq.~(\ref{eq:c})).
We also ignored the $q_x$ dependence of the symmetric term of the interaction matrix, since it gives rise only to a sub-leading correction.
The integral in Eq.~(\ref{eq:unscrgreen}) can be evaluated exactly by changing variables to polar coordinates to get
\be
	{\cal C} (\tau ) = -\frac{8}{3}\frac{\Gamma (-\frac{2}{3})}{\Gamma (\frac{2}{3})}(\frac{V_C^2}{Yl^2}\tau )^{1/3}.
\ee
The prefactor $-\frac{8}{3} \Gamma (-\frac{2}{3})/\Gamma (\frac{2}{3}) \approx 8$.
The time scale $\tau_0 = (Yl^2)/V_C^2$ is very small.  For example, evaluating $Y$ using Eq.~(\ref{eq:y}) for $L = 2$, $\nu^* = 0.4$, we find that $\tau_0$ approximately equals $0.006/V_C$ and is still smaller for higher Landau levels (for $L = 3$, $\tau_0 = 0.0038/V_C$; for $L = 6$, $\tau_0 = 0.003/V_C$, etc.).

Similarly, for screened Coulomb interactions, we have 
\be
	{\cal C}(\tau ) =  \frac{2}{\pi l^2}\sqrt{\frac{V(0)+\chi^{-1}}{Y}} \int \frac{d{\bf q}}{|q_xq_y|} (1-e^{-|q_xq_y|l^2 \Omega_0\tau }),\label{eq:scrgreen}
\ee
where $\Omega_0 = \frac{1}{2}\sqrt{Y(V(0)+\chi^{-1})}$.
In principle, the integral in Eq.~(\ref{eq:scrgreen}) can be expressed exactly in terms of a generalized hypergeometric function, ${\cal C}(\tau ) =$ $ \frac{2}{\pi l^2}\sqrt{\frac{V(0)+\chi^{-1}}{Y}} {}_3F_3(\{ 1,1,1 \},\{ 2,2,2 \}, -\pi^2l/a \Omega_0 \tau )$, however, an approximate but intuitive form may be more useful.
To analyze the leading behavior of the integral as a function of $\tau $, we split the integration domain into two:  one in which $|q_xq_y|l^2 < 1/(\Omega_0 \tau) $ (where $\Omega_0 = 1/2\sqrt{Y(V(0)+\chi^{-1})}$) and the exponential function in the integrand can be expanded into a Taylor series, and another, in which $|q_xq_y|l^2 > 1/(\Omega_0 \tau ) $ and we assume that the exponential is much smaller than 1 and can be ignored.  
The leading behavior comes from the boundaries of the domain, in which $|q_xq_y|l^2 > 1/(\Omega_0 \tau ) $, and we obtain
\be
	{\cal C}(\tau ) \approx \frac{1}{\pi l^2}\sqrt{\frac{V(0)+\chi^{-1}}{Y}}\log^2(\pi^2\frac{l}{a}\Omega_0\tau ).
\ee
Here the time-scale $1/(\pi^2\frac{l}{a}\Omega_0)$ is proportional to $1/V_C$ with a proportionality constant of order unity.

It is interesting to compare these results to the tunneling exponents of a fractional quantum Hall state at half-filling.  
Unlike the striped state, the fractional quantum Hall state is isotropic, but its upper (or only) Landau level is also partially filled.  
As one would expect, tunneling into the isotropic state is more suppressed than that into the striped state:\cite{halperin,wen} ${\cal C} (\tau )$ behaves as $\tau^{1/2}$ for unscreened Coulomb interactions and $\tau^{1/3}$ for screened Coulomb interactions.
The results are summarized in Table \ref{tab:res}.

The low frequency behavior of the spectral function, which can be obtained by taking the inverse Laplace transform of $G_e(\tau )$, gives the behavior of the current-voltage relations (I-V curves).\cite{halperin,wen}
The results for I(V) are summarized in Table \ref{tab:res2}.

\section{Further Applications and Conclusions}

The Luttinger liquid model and its parameters derived in this paper can be applied to other calculations and studies of various properties of the striped quantum Hall state.
Thus, for example, Wexler and Dorsey use the Luttinger liquid parameters in the long wavelength limit to study disclination unbinding\cite{wexler1}.  They use the parameters as a starting point in their RG analysis and find the Kosterlitz-Thouless temperature for the transition between orientationally ordered and isotropic phases of the striped state.

MF use the Luttinger liquid interaction matrix at $q_x=0$ to study the stability of the striped state with respect to backscattering\cite{macdonald1}.  
They add a backscattering term to the Luttinger liquid Lagrangian and analyze the flow of the backscattering amplitudes when $\omega $ and $q_x$ are rescaled (while $q_y$ serves as a fixed parameter).
Their RG analysis shows that backscattering is always relevant, and, at very low temperatures ($<$10mK), the striped state is unstable with respect to formation of an anisotropic Wigner crystal.  
We have reproduced their calculation and found that, despite some definitional differences, our result for the scaling dimension of the backscattering term is identical to that of MF. 

We also considered the scaling of the backscattering term in the presence and absence of rotational symmetry, since in a real sample there are always preferred directions because of the geometry of the sample and the pinning of the stripes by impurities.  
The assumption of rotational symmetry manifests itself in the form of the smectic elasticity $D_S({\bf q}) \propto Yq_y^2 +K q_x^4$.  
In the absence of rotational symmetry the quadratic term $q_x^2$ in the elasticity does not vanish and, in the long wavelength limit, the elasticity takes the form: $D_S({\bf q}) \propto Yq_y^2 +K_2 q_x^2$.  
The RG analysis can depend on the presence or absence of rotational symmetry if one chooses to rescale both $q_x$ and $q_y$ (as opposed to treating $q_y$ as a fixed parameter).   
However, for short-range interactions, the scaling dimension of the backscattering term is independent of rotational symmetry, {\it i.e.} the absence of rotational symmetry does not increase the stability of the striped state.   
It is important to note, however, that the rotational symmetry might make a difference in stability of the striped state with long-range interactions.

A much more detailed RG analysis of the effects of disorder is given in Ref.~\onlinecite{scheidl}.  
In their paper, Scheidl and von Oppen use an elastic theory of the striped states, which is very similar to the Luttinger liquid theory, to map out a phase-diagram of different disorder and length-scale regimes\cite{scheidl}. 

The hydrodynamic model, for which we found all the parameters by establishing a one-to-one correspondence with the Luttinger liquid theory, has been introduced and used by FV in Ref.~\onlinecite{fogler2} to find dynamical scaling exponents of the striped state,

In addition to these applications and the applications described in the previous two sections, the model and the parameters will probably find a wide range of other uses in the analysis of the striped quantum Hall states.  

In summary, we have derived an effective interacting Luttinger liquid model for the low energy excitations of the striped quantum Hall state on the basis of a first-principles Hartree-Fock calculation. 
Using our coherent state formulation we were able to describe all possible low energy excitations (with low $q_x$ and any $q_y$) of the striped state, and thus derived the full interaction matrix for the Luttinger liquid model, complete with all parameters.

In the limit of long wavelengths, we have established a one-to-one correspondence between the Luttinger liquid model and the hydrodynamic model of Ref.~\onlinecite{fogler2}.  
Thus, we have also obtained expressions for the parameters of the hydrodynamic model from Hartree-Fock.

We have also given some examples of applications of the Luttinger liquid model and its parameters.  
In Sec.~\ref{sec:dynamics} we have found the dispersion relation for the excitations of the striped state and in Sec.~\ref{sec:tunneling} we have discussed the behavior of perpendicular tunneling into the striped state and obtained the corresponding I-V relations.  
Further applications of our results are almost certain to be found.

\section*{Acknowledgements}

This work was supported in part by the NSF through the GRF program (A.L.), the MRSEC program DMR98--08941 (X.-G.W.), under grant DMR97--14198 (X.-G.W.), and under grant DMR99--81283 (B.I.H.).  
A.L. would like to thank Lucent Technologies Bell Labs for hospitality and support under the GRPW program.

\begin{table*}[h]
\begin{tabular}{|l|c|c|}
	           & $\nu^* \approx 1/2, L > 1$ & $\nu = 1/2, L=0$ \\
		   & striped            & isotropic \\ \hline
	unscreened & $\tau^{1/3}$ & $\tau^{1/2}$ \\ \hline
	screened   & $\log^2 (\Omega_0 \tau)$ & $\tau^{1/3}$ 
\end{tabular}
	\caption{\label{tab:res} Large-$\tau $ 
behavior of ${\cal C}(\tau )$ (${\cal C}(\tau ) = -\log G_e(\tau )$, where $G_e(\tau )$ is the electron Green's function) for the striped state and isotropic state of a 2DEG with a half-filled upper Landau level (the results for the isotropic case are from Refs.~21 and 22). 
 For both unscreened and screened Coulomb interaction, tunneling is weaker in the isotropic case.}
\end{table*}

\begin{table*}[h]
\begin{tabular}{|l|c|c|}
	           & $\nu^* \approx 1/2, L > 1$ & $\nu \approx 1/2, L=0$ \\
		   & striped            & isotropic \\ \hline
	unscreened & $ \exp(-\mbox{const}/\sqrt{V}) $ & $\exp(-\mbox{const}/V) $ \\ \hline
	screened   & $ \exp(-\mbox{const}\log^2 (V))$ & $\exp(-\mbox{const}/\sqrt{V}) $ 
\end{tabular}
	\caption{\label{tab:res2} General behavior of the I-V curves for the striped state and isotropic state of a 2DEG with a half-filled upper Landau level (the results for the isotropic case are from 
Refs.~21 and 22)
.}
\end{table*}

\end{multicols}


\begin{references}
	\bibitem{fukuyama} H.~Fukuyama, P.~M.~Platzman, and P.~W.~Anderson, Phys~Rev.~B {\bf 19} 5211 (1979).

	\bibitem{shayegan} See for example, M.~Shayegan in {\it Perspectives in Quantum Hall Effects}, ed.~S.~Das Sarma and A.~Pinczuk (Wiley, 1997).

	\bibitem{experiment1} M.~P.~Lilly, K.~B.~Cooper, J.~P.~Eisenstein, L.~N.~Pfeiffer, and K.~W.~West, Phys.~Rev.~Lett. {\bf 82}, 394 (1999).

	\bibitem{stormer} R.~R.~Du, D.~C.~Tsui, H.~L.~Stormer, L.~N.~Pfeiffer, K.~W.~Baldwin, and K.~W.~West, Solid State Comm.~{\bf 109}, 389 (1999).

	\bibitem{comment} The first observations of unexplained behavior in high Landau Levels -- which has now been identified with anisotropic striped states in retrospect -- were probably by R.~L.~Willet and J.~P.~Eisenstein, 1988 (unpublished).  The first published report of these behaviors was by H.~L.~Stormer {\it et al.} Bull.~Am.~Phys.~Soc.~{\bf 38}, 235 (1993).

	\bibitem{fogler1} A.~A.~Koulakov, M.~M.~Fogler, and B.~I.~Shklovskii,
  Phys.~Rev.~Lett. {\bf 76}, 499 (1996); Phys.~Rev.~B {\bf 54}, 1853
  (1996). 

	\bibitem{moessner} R.~Moessner and J.~T.~Chalker, Phys.~Rev.~B {\bf 54}, 5006 (1996).

	\bibitem{macdonald1} A.~H.~MacDonald and M.~P.~A.~Fisher, Phys.~Rev.~B {\bf 61}, 5724 (2000).

	\bibitem{fertig1} R.~C\^{o}t\'{e} and H.~A.~Fertig, Phys.~Rev.~B {\bf 62}, 1993 (2000).

	\bibitem{haldane} E.~H.~Rezayi, F.~D.~M.~Haldane, Kun Yang, Phys.~Rev.~Lett.~{\bf 83}, 1219 (1999).

	\bibitem{fogler2} M.~M.~Fogler and V.~M.~Vinokur, Phys.~Rev.~Lett.~{\bf 84}, 5828 (2000).

	\bibitem{wexler1} C.~Wexler and A.~T.~Dorsey, cond-mat/0009096.

	\bibitem{scheidl} S.~Scheidl and F.~von Oppen, cond-mat/0007442.

	\bibitem{aleiner1} I.~L.~Aleiner and L.~I.~Glazman, Phys.~Rev.~B {\bf 60}, 11296 (1995), and Ref.~9 therein.

	\bibitem{fourier} The Fourier conventions used in the paper:
		\bea
			f_{{\bf q}\alpha } & = & a \sum_I \int_x e^{-iq_xx-iq_yaI} f_{I\alpha }(x) \nonumber\\
			f_{I\alpha }(x) & = & \int_{-\infty}^{\infty} \frac{dq_x}{2\pi }\int_{-\pi /a}^{\pi /a}\frac{dq_y}{2\pi }e^{iq_xx+iq_yaI} f_{{\bf q}\alpha }. \nonumber
		\eea

	\bibitem{fradkin1} E.~Fradkin and S.~A.~Kivelson, Phys.~Rev.~B {\bf 59}, 8065 (1999).

	\bibitem{macdonalds-y} The quantity $K_y$ calculated in Ref.~7 is not the same as $Y$ but is related to $Y$ as $K_y = l^2a^2Y-\frac{ma}{\pi } C a\nu^* + \frac{1}{2\pi }\frac{\chi ^{-1}}{2 \pi l^2}(a\nu^*)^2$.  See also Sec.~IIIB. 

	\bibitem{chaikin} P.~M.~Chaikin and T.~C.~Lubensky, {\it Principles of Condensed Matter Physics}, p.~311 (Cambridge University Press, 1995).

	\bibitem{macdonald2} T.~Jungwirth, A.~H.~MacDonald, L.~Smr\v{c}ka, and S.~M.~Girvin, Phys.~Rev.~B {\bf 60}, 15574 (1999).

	\bibitem{gradshteyn} I.~S.~Gradshteyn and I.~M.~Ryzhik, {\it Table of Integrals, Series, and Products} (Academic Press, 1994).

	\bibitem{halperin} S.~He, P.~M.~Platzman, and B.~I.~Halperin, Phys.~Rev.~Lett.~{\bf 71}, 777 (1993).

	\bibitem{wen} Y.~B.~Kim and X.-G.~Wen, Phys.~Rev.~B {\bf 50}, 8078 (1994).


\end{references}
\end{document}